\documentclass[usenatbib]{mn2e}

\def\lapp{\ifmmode\stackrel{<}{_{\sim}}\else$\stackrel{<}{_{\sim}}$\fi}
\def\gapp{\ifmmode\stackrel{>}{_{\sim}}\else$\stackrel{>}{_{\sim}}$\fi}

\usepackage{xspace}
\usepackage{natbib}
\usepackage{graphicx}
\usepackage{float}
\usepackage{ulem}
\usepackage[T1]{fontenc}
\usepackage{lmodern}

\newcommand{\presto}{\texttt{PRESTO}\xspace}

\newcommand{\prepfold}{\texttt{prepfold}\xspace}
\newcommand{\psrchive}{\texttt{psrchive}\xspace}
\newcommand{\pat}{\texttt{pat}\xspace}
\newcommand{\pam}{\texttt{pam}\xspace}
\newcommand{\nemodel}{\texttt{NE2001}\xspace}

\newcommand{\tempotwo}{\texttt{TEMPO2}\xspace}

\newcommand{\psr}{PSR~J1952+2630\xspace}

\def\apjl{ApJL}
\def\apj{ApJ}
\def\mnras{MNRAS}
\def\aap{A\&A}
\def\nat{Nat.}

\def\pasp{PASP}
\def\apjs{ApJ Supp.}
\def\aj{AJ}

\def\physrep{Phys. Rep.}

\def\pasj{PASJ}
\def\aapr{A\&A Rev.}
\def\prd{Phys.~Rev.~D}
\def\pasa{PASA}               % Publications of the Astron. Soc. of Australia

\title[Young Recycled PSR with a Massive WD companion]
{Timing of a Young Mildly Recycled Pulsar with a Massive White Dwarf Companion}

\author[P.~Lazarus et al.]
{P.~Lazarus$^{1}$\thanks{E-mail: plazarus@mpifr-bonn.mpg.de},
T.~M.~Tauris$^{2,1}$,
B.~Knispel$^{3,4}$,
P.~C.~C.~Freire$^{1}$,
\newauthor
J.~S.~Deneva$^{5}$,
V.~M.~Kaspi$^{6}$,
B.~Allen$^{3,4,7}$,
S.~Bogdanov$^{8}$,
\newauthor
S.~Chatterjee$^{9}$,
I.~H.~Stairs$^{10}$,
and W.~W.~Zhu$^{10}$
\\ % The newline mark-up must be on its own line or else the tex docuemnt doesn't compile!
$^1$Max-Planck-Institut f\"ur Radioastronomie, Auf dem H\"ugel 69, 53121 Bonn, Germany\\
$^2$Argelander-Institut f\"ur Astronomie, Universit\"at Bonn, Auf dem H\"ugel 71, 53121 Bonn, Germany\\
$^3$Max-Planck-Institut f\"ur Gravitationsphysik, D-30167 Hannover, Germany\\
$^4$Leibniz Universit{\"a}t Hannover, D-30167 Hannover, Germany\\
$^5$Naval Research Laboratory, 4555 Overlook Ave SW, Washington, DC 20375\\
$^6$Dept.~of Physics, McGill Univ., Montreal, QC H3A 2T8, Canada\\
$^7$Physics Dept., U. of Wisconsin - Milwaukee, Milwaukee WI 53211, USA\\
$^8$Columbia Astrophysics Laboratory, Columbia Univ., New York, NY 10027, USA\\
$^9$Astronomy Dept., Cornell Univ., Ithaca, NY 14853, USA\\
$^{10}$Dept.~of Physics and Astronomy, Univ.~of British Columbia, Vancouver, BC V6T 1Z1, Canada
}

\begin{document}
\maketitle

\begin{abstract}
We report on timing observations of the recently discovered binary pulsar \psr
using the Arecibo Observatory. The mildly recycled 20.7-ms pulsar is in a
9.4-hr orbit with a massive, $M_{WD} > 0.93\;M_{\odot}$, white dwarf (WD)
companion. We present, for the first time, a phase-coherent timing solution,
with precise spin, astrometric, and Keplerian orbital parameters. This shows
that the characteristic age of \psr is $77\;{\rm Myr}$, younger by one order of
magnitude than any other recycled pulsar--massive WD system. We derive an upper
limit on the true age of the system of $150\;{\rm Myr}$. We investigate the
formation of \psr using detailed modelling of the mass-transfer process from a
naked helium star on to the neutron star following a common-envelope phase
(Case BB Roche-lobe overflow). From our modelling of the progenitor system, we
constrain the accretion efficiency of the neutron star, which suggests a value
between 100 and 300\% of the Eddington accretion limit. We present numerical
models of the chemical structure of a possible oxygen-neon-magnesium WD
companion.  Furthermore, we calculate the past and the future spin evolution of
\psr, until the system merges in about 3.4~Gyr due to gravitational wave
emission.  Although we detect no relativistic effects in our timing analysis we
show that several such effects will become measurable with continued
observations over the next 10 years; thus \psr has potential as a testbed for
gravitational theories.
\end{abstract}


\section{Introduction}
\label{sec:intro}

Since 2004, the Arecibo L-band Feed Array (ALFA), a 7-beam receiver at the
focus of the 305-m William E. Gordon radio telescope at the Arecibo
Observatory, is being used to carry out the Pulsar--ALFA (PALFA) survey, a deep pulsar survey
of low Galactic latitudes \citep{cfl+06,laz13,nab+13}.
% cfl+06: http://adsabs.harvard.edu/abs/2006ApJ...637..446C
% laz13: http://adsabs.harvard.edu/abs/2013IAUS..291...35L
Given its short pointings, the PALFA survey is especially sensitive to binary
pulsars in tight orbits, as demonstrated by the discovery of the relativistic binary pulsar PSR J1906+0746,
which did not require any acceleration search techniques \citep{lsf+06}.
% lsf+06: http://adsabs.harvard.edu/abs/2006ApJ...640..428L
Another aspect of this and other modern Galactic plane surveys is the high time
and frequency resolution, which allow the detection of millisecond pulsars (MSPs) at
high dispersion measures \citep{crl+08,dfc+12,csl+12}
% crl+08: http://adsabs.harvard.edu/abs/2008Sci...320.1309C
% dfc+12: http://adsabs.harvard.edu/abs/2012ApJ...757...89D
% csl+12: http://adsabs.harvard.edu/abs/2012ApJ...757...90C
and therefore greatly expand the volume in which these can be discovered.

One of the innovative aspects of this survey is the use of distributed,
volunteer computing. One of the main motivations is the detection of extremely
tight (down to $P_b \sim 10$ minutes) binaries, for which acceleration and jerk
searches become computationally challenging tasks. The analysis of survey data is
distributed through the Einstein@Home (E@H) infrastructure
\citep{kac+10,akc+13}.
% kac+10: http://adsabs.harvard.edu/abs/2010Sci...329.1305K
% akc+13: http://arxiv.org/abs/1303.0028
Thus far, the E@H pipeline has discovered 24 new pulsars in the PALFA survey
data alone, complementing the other data analysis pipelines the PALFA survey
employs (Quicklook, and PRESTO; Stovall et. al, in prep., and Lazarus et. al, in prep., respectively).
%As well as, 11 new pulsars
%in gamma-ray data \citep{pga+12a,pga+12b,pgf+12}
% pga+12a: http://adsabs.harvard.edu/abs/2012ApJ...744..105P
% pga+12b: http://adsabs.harvard.edu/abs/2012ApJ...755L..20P
% pgf+12: http://adsabs.harvard.edu/abs/2012Sci...338.1314P
%and 24 more in Parkes Multibeam archival data \citep{kek+13}.
% kek+13: http://arxiv.org/abs/1302.0467

\psr was the first binary pulsar discovered with the E@H pipeline
\citep{kla+11}. At that time the few observations available allowed
only a rough estimate of the orbital parameters of this MSP based on Doppler
measurements of the spin period. These already showed that \psr has a
massive WD companion ($M_{WD} > 0.945 {\rm M_\odot}$ assuming $M_p = 1.4 {\rm
M_\odot}$), and may have evolved from an intermediate-mass X-ray
binary (IMXB). Building on the analysis by \citet{kla+11}, we present
in this paper the phase-coherent timing solution of \psr resulting from
dedicated follow-up observations with the Arecibo telescope, which provides
orbital parameters far more precise than those previously determined. Our
timing solution also shows the system is relatively young ($\tau_c$ = 77~Myr).

It is commonly accepted that MSPs are spun~up to their high spin
frequencies via accretion of mass and angular momentum from a companion star
\citep{acrs82,rs82,bh91}.  In this recycling phase the system is observable as
an X-ray binary \citep[e.g.][]{hay85,nag89,bcc+97} and towards the end of
this phase as an X-ray MSP \citep{wv98,pfb+13}.  

The majority of MSPs have helium WD companions and their formation is mainly
channeled through low-mass X-ray binaries (LMXBs) which have been well
investigated in previous studies
\citep[e.g.][]{wrs83,ps88,ps89,rpj+95,esa98,ts99,prp02,ndm04,vvp05}.  In contrast, binary
pulsars, such as \psr, with relatively heavy WDs ({CO} or {ONeMg} WDs) are less common in
nature. Their formation and recycling process involves a more massive WD
progenitor star in an IMXB \citep[see][and references therein, for a discussion
of their suggested formation channels]{tlk11}. Here we distinguish IMXBs from
other X-ray binaries as systems that leave behind a massive WD companion rather
than a neutron star (NS). Some of these IMXB systems with donor stars of
$6-7\;M_{\odot}$ could also be classified observationally as Be/X-ray binaries since these
stars are of spectral class B3-4 with emission lines. Recently, \citet{tlk12}
presented a detailed study of the recycling process of pulsars via both LMXBs
and IMXBs and highlighted their similarities and differences. These authors
also presented the first calculations of mild recycling in post common envelope
systems where mass transfer proceeds via so-called Case~BB Roche-lobe overflow
(RLO; see Sec.~\ref{sec:evolution} for details).

\psr's combination of a young, massive WD in a close orbit with a recycled pulsar
poses interesting questions about its formation and future evolution.  Binary
MSPs represent the advanced stage of stellar evolution in close,
interacting binaries. Their observed orbital and stellar properties are thus
fossil records of their evolutionary history. Therefore by using the precise
description of a pulsar binary system determined from phase-coherent timing,
and binary evolution modelling, we use \psr as a probe of
stellar astrophysics.

We also demonstrate that \psr is an interesting test case
for Case~BB RLO, enabling interesting constraints on the accretion physics from
the combined modelling of binary stellar evolution and the spin kinematics of
this young, mildly recycled pulsar.
\\

The rest of this paper is presented as follows: Sec.~\ref{sec:observations}
describes the observations of \psr,  and details of the data reduction and
timing analysis. Results from this analysis are presented in
Sec.~\ref{sec:results}. The binary evolution of the system is detailed in
Sec.~\ref{sec:evolution}. The implications of our results, and future prospects
are described in Secs.  \ref{sec:discussion}, and \ref{sec:future},
respectively. Finally Sec.~\ref{sec:conclusion} summarizes the paper.

\section{Observations and Data Analysis}
\label{sec:observations}

Following its discovery in 2010 July \psr was observed during PALFA survey
observing sessions using the usual survey observing set-up: the 7-beam ALFA
receiver with the Mock
spectrometers\footnote{http://www.naic.edu/$\sim$astro/mock.shtml}. In this
set-up, $\sim$322~MHz ALFA observing band was split into two overlapping
sub-bands centred at 1300.1680~MHz and 1450.1680~MHz, each with a bandwidth of
172.0625~MHz. All timing observations using this set-up were performed with
ALFA's central beam, and were typically 5-10 minutes in duration.

A dedicated timing program at the Arecibo Observatory started in 2011
November. The dedicated timing observations took data using the ``L-wide'' receiver. These
data were divided into four non-overlapping, contiguous sub-bands recorded by the Mock
spectrometers in search-mode. Each sub-band has 172~MHz of bandwidth divided in
2048 channels, sampled every $\sim$83.3~$\mu$s. Together the four sub-bands
cover slightly more than the maximum bandwidth of the receiver,
580~MHz, and are centred at 1444~MHz. 

At the start of the dedicated timing campaign, three 3-hour observing sessions
on consecutive days were conducted to obtain nearly complete orbital coverage,
with the goal of detecting, or constraining a Shapiro delay signature caused by
the pulsar's signal passing through its companions gravitational potential
well.

Subsequently, monthly observations of 1~hour each were used for the next
11~months to monitor the pulsar, and refine our timing solution.

%\section{Data Analysis}
%\label{sec:analysis}

For analysis, each observation was divided into segments no longer than
15-minutes for each of the separate sub-bands. Each segment was folded offline
using the appropriate topocentric spin period using \prepfold program from
\presto\footnote{http://www.cv.nrao.edu/$\sim$sransom/presto/}. The resulting
folded data files were converted to PSRFITS format \citep{hsm04} with
\pam from the \psrchive suite of pulsar analysis
tools\footnote{http://psrchive.sourceforge.net/}. Data files were fully time
and frequency integrated, and times-of-arrival (TOAs) were computed
using \pat from \psrchive. The result was a single TOA for each $\sim15$ minutes
of observing for each sub-band.

TOAs for the different sub-bands were computed using a separate analytic template
produced by fitting von Mises functions to the sum of profiles from the
closely spaced three-day observing campaign. Profiles from different observations
were aligned using the pulsar ephemeris. 
%Templates are shown in Fig.~\ref{fig:profs}.

%\begin{figure}
%    \includegraphics[width=\columnwidth]{figures/template/J1952+2630_profiles.ps}
%    \caption{Analytic templates used for timing. Each template is determined
%    from fitting von Mises functions to the sum of all data for a single
%    sub-band from observing sessions on three consecutive days.
%    \label{fig:profs}}
%\end{figure}

\begin{figure}
    \includegraphics[width=\columnwidth]{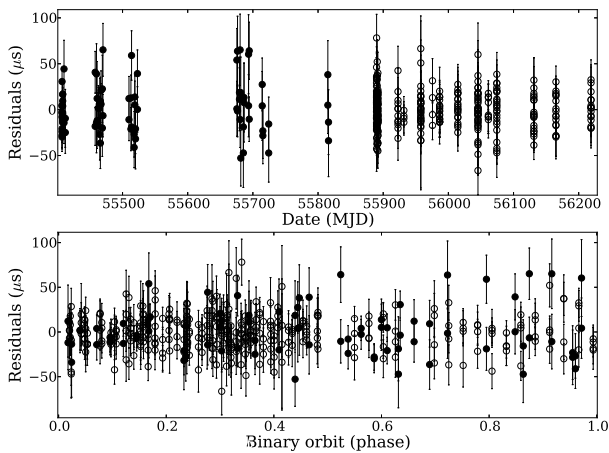}
    \caption{The difference between our pulsar TOAs and our timing
    solution. The filled circles are TOAs from data taken with the ALFA
    receiver, and the un-filled circles TOAs from data taken with the L-wide
    receiver. No systematic trends are visible as a function of epoch, or
    orbital phase.
    \label{fig:residuals}}
\end{figure}

Timing analysis was performed with the \tempotwo software \citep{hem06}. The
phase-coherent timing solution determined fits the 418 pulse times-of-arrival.
Our timing solution accurately models the timing data, leaving no
systematic trends as a function of epoch, or orbital phase (see
Fig.~\ref{fig:residuals}).

\section{Results}
\label{sec:results}

The timing analysis of \psr has resulted in the determination of astrometric, spin, and
Keplerian orbital parameters. The fitted timing parameters, as well as some
derived parameters can be found in Table~\ref{tab:timing}. The timing solution
also includes a marginal detection of the proper motion of the pulsar. 

\subsection{The Nature of the Binary Companion Star}
\label{subsec:nature} 

Given the combination of a relatively large observed mass function,
$f=0.153\;M_{\odot}$ (see Table~\ref{tab:timing}) and a small orbital
eccentricity, $e=4.1\times 10^{-5}$, it is clear that \psr has a massive WD
companion.  The minimum companion mass is $M_{\rm WD}^{\rm min}\approx
0.93\;M_{\odot}$, obtained for an orbital inclination angle of $i=90^{\circ}$
and an assumed NS mass, $M_{\rm NS}=1.35\;M_{\odot}$.  The small eccentricity excludes
a NS companion star since the release of the gravitational binding energy
alone, during the core collapse, would make the post-SN eccentricity much
larger \citep[$e \gg 0.01$;][]{bh91}, in contrast with the observed
value. Furthermore, no known double-neutron star system has $e < 0.01$, according
to the ATNF Pulsar Catalogue\footnote{http://www.atnf.csiro.au/people/pulsar/psrcat/}
\citep{mhth05}. Thus a double NS system is not possible and \psr must have a
massive WD companion, i.e. a carbon-oxygen (CO) or an oxygen-neon-magnesium
(ONeMg)~WD.  The upper (Chandrasekhar) mass limit for a rigidly rotating WD is
$\sim\!1.48\;M_{\odot}$ \citep[e.g.][]{yl05} and therefore we conclude that the
WD companion star is in the mass interval $0.93 \lapp M_{\rm WD}/M_{\odot} \lapp 1.48$. 

The distance to \psr, estimated using the observed dispersion measure, and the
\nemodel model of Galactic free electrons, is $d \simeq 9.6\;{\rm kpc}$
\citep{cl02}. The uncertainty in DM-derived distances using the \nemodel model
can, in some cases, be up to a factor of $\sim2$ off from the true distance.
Unfortunately, this places the binary system too far away to hope to optically detect the
WD companion with current telescopes. As expected, a search of optical and infrared
catalogs yielded no counterpart.

\subsection{The age of \psr}
\label{subsec:age} 

\psr has a spin period of $P=20.7\;{\rm ms}$ and one of the highest values of
the spin period derivative, $\dot{P}=4.27\times 10^{-18} \rm s\,s^{-1}$ for any known
recycled pulsar\footnote{This comparison was made using the ATNF Pulsar
Catalogue.} and by far the highest value for a mildly recycled pulsar with a
massive WD companion.  The observed value of $\dot{P}$ is contaminated
by kinematic effects (see Sec.~\ref{sec:future}). However, the contamination is
only $\sim$0.01\%, assuming our current value of the proper motion. The
combination of $P$ and $\dot{P}$ of \psr yields a small characteristic age,
$\tau\equiv P/2\dot{P} \simeq 77\;{\rm Myr}$.  The characteristic age of a
pulsar should only be considered a rough order-of-magnitude estimate of the
true age of the pulsar (i.e. time since recycling terminated). Thus the
\textit{true} ages are quite uncertain for recycled pulsars with large
$\tau$-values of several Gyr \citep{tau12,tlk12}, unless a cooling age of their
WD companion can be determined. However, the true ages of recycled pulsars with
small values of $\tau $ (less than a few $100\;{\rm Myr}$) are relatively close
to the characteristic age.  Hence, we conclude that \psr is young, for a
recycled pulsar, and in Secs~\ref{subsec:age-spin} and
\ref{subsubsec:isochrones} we discuss its true age (i.e. its actual age since
it switched-on as a recycled radio pulsar) and also constrain its spin
evolution in the past and in the future.

\subsection{Constraints on the Binary System from Timing}
\label{subsec:timingconstraints}

Given the current timing data, the binary motion of \psr can be accurately
modelled without requiring any relativistic, post-Keplerian parameters.
Unfortunately, this means that the masses of the pulsar and its WD companion
cannot be precisely determined by the current timing model.

\begin{figure}
    \includegraphics[width=\columnwidth]{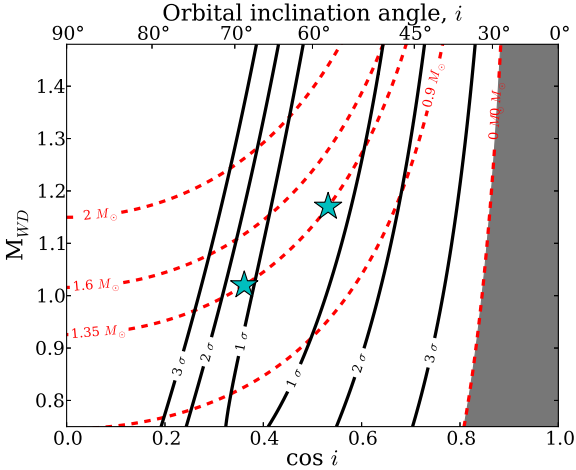}
    \caption{Map of companion-mass and inclination combinations allowed by the
    current timing data. The 1-,2-, and 3-$\sigma$ contours, shown in black,
    enclose $\sim$68.3, $\sim$95.5, and $\sim$99.7~\% of the allowed binary
    system configurations, respectively, given the current timing data, and
    requiring the companion's mass to lie in the range 
    $0.75 \leq M_{\rm WD} \leq 1.48\;M_{\odot}$.
    The red dashed lines trace constant pulsar mass. The grey region in the
    right is excluded because the pulsar mass must be larger than 0. The
    two stars are the results from the two simulations of the binary system's
    evolution (see Sec.~\ref{subsec:CaseBB}).
    \label{fig:m2cosi}}
\end{figure}

Nevertheless, a $\chi^2$ analysis was performed to investigate what constraints
the lack of a Shapiro delay detection imposes on the make-up and geometry of the
binary system (i.e. the mass of the pulsar, $M_{\rm NS}$, and that of its
companion, $M_{\rm WD}$, as well as the binary system inclination, $i$, by
using the mass-function). 
A $\chi^2$-map was computed using the \texttt{ChiSqCube} plug-in\footnote{The
\texttt{ChiSqCube} plug-in populates a 3-dimensional $\chi^2$ space. For the
purpose of the analysis presented here, the third dimension was not used.} for
\tempotwo (see Fig.~\ref{fig:m2cosi}).  The $\chi^2$-map was computed for
$M_{\rm WD}$ between 0 and 1.5 $M_\odot$, in steps of 0.001 $M_\odot$, and $\cos i$
from 0 to 1 in steps of 0.001.

For each point in the $\chi^2$-map the timing model was re-fit, holding the
values of $M_{\rm WD}$ and $\cos i$ fixed. The resulting $\chi^2$ values were
converted to probabilities by following \citet{sna+02}, and then normalized.
The 1-, 2-, and 3-$\sigma$ contours were chosen such that they contain
$\sim$68.3, $\sim$95.5, and $\sim$99.7~\% of the allowed binary system
configurations. Based on this analysis we know the binary system cannot be
edge-on ($i = 90^{\circ}$). The inclination angle is constrained to be be $i
\leq 75^{\circ}$ for $M_{\rm NS} \geq 1.35 M_\odot$, at the 3-$\sigma$ level.

However, based on the massive companion, and small, but significantly non-zero
eccentricity, we expect that with our current timing precision we will measure two (or
possibly three) post-Keplerian parameters precisely enough to provide a
stringent test of relativistic gravity, within the next 10 years (see Sec.~\ref{sec:future}).

\section{Binary Evolution of the Progenitor} 
\label{sec:evolution}

\begin{figure} 
    \centering
    \includegraphics[width=1.05\columnwidth,angle=0]{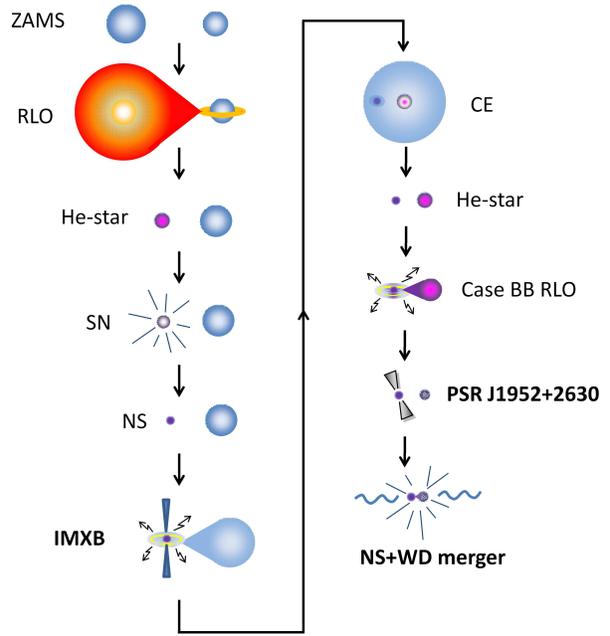} 
    \caption{An illustration of the full binary stellar evolution from the zero-age main
        sequence (ZAMS) to the final merger stage.  The initially more massive star
        evolves to initiate Roche-lobe overflow (RLO), leaving behind a naked helium
        core which collapses into a neutron star (NS) remnant, following a supernova
        explosion (SN).  Thereafter, the system becomes a wide-orbit intermediate-mass X-ray
        binary (IMXB), leading to dynamically unstable mass transfer and the formation
        of a common envelope (CE), when the 6--7$\;M_{\odot}$ donor star initiates RLO.
        The post-CE evolution, calculated in detail in this paper, is responsible for
        recycling the NS via Case~BB RLO when the helium star companion expands to
        initiate a final mass-transfer episode.  \psr is currently observed as a
        mildly recycled radio pulsar orbiting a massive white~dwarf (WD).  The system
        will merge in $\sim3.4$~Gyr, possibly leading to a $\gamma$-ray burst-like event
        and the formation of a single black hole. 
        \label{fig:cartoon}} 
\end{figure}

In general, binary pulsars with a CO~WD companion\footnote{Here, and in the
following, we simply write `CO~WD' for any massive WD whose exact chemical
composition (CO or ONeMg) is unknown.} can form via different
formation channels \citep[see][and references therein]{tlk12}.

Pulsars with CO WDs in orbits of $P_{\rm orb}\le$~2--3~days, like \psr
($P_{\rm orb}=9.4$ hr), are believed to have formed via a common
envelope (CE) scenario. Such systems originate from IMXBs which have very large
values of $P_{\rm orb}$ prior to the onset of the mass transfer.  These
systems are characterized by donor stars with masses between
$2<M_2/M_{\odot}<7$, and very wide orbits up to $P_{\rm orb}\simeq 10^3$~days.
Donor stars near the tip of the red giant branch or on the asymptotic giant
branch evolve via
late Case~B RLO or Case~C RLO, respectively.  As a result, these
donor stars develop a deep convective envelope as they enter the giant phase,
before filling their Roche~lobe.  These stars respond to mass loss by
expanding, which causes them to overfill their Roche~lobe even more.  Binaries
where mass transfer occurs from a more massive donor star to a less massive
accreting NS shrink in size, causing further overfilling of the donor star's
Roche~lobe, which further enhances mass loss. This process leads to a
dynamically unstable, runaway mass transfer and the formation of a CE
\citep{pac76,il93,ijc+12}.  However, the wide orbit prior to the RLO is also
the reason why these systems survive the CE and spiral-in phase, since the
binding energy of donor star's envelope becomes weaker with advanced stellar
age, and therefore the envelope is easier to eject, thereby avoiding a merger
event.

%For discussions on so-called early Case~B RLO of IMXBs, which may also lead to
%the formation of binary pulsars with CO~WD companions, see \citet{tvs00} and
%\citet{pr00}. For detailed calculations of Case~A RLO, see e.g. \citet{lrp+11}
%and \citet{tlk11} who studied the formation of PSR~J1640$-$2230 -- the only
%known fully recycled millisecond pulsar with a CO~WD companion \citep{dpr+10}.

Given that the duration of the CE and spiral-in phase is quite short
\citep[$<10^3\;{\rm yr}$; e.g.][]{pod01,pdf+12,ijc+12} the NS can only accrete
$\sim\!10^{-5}\;M_{\odot}$ during this phase, assuming that its
accretion is limited by the Eddington accretion rate (a few
$10^{-8}\;M_{\odot}\,{\rm yr}^{-1}$). This small amount is not enough to even
mildly recycle the pulsar. Instead, the NS is thought to be recycled during the
subsequent so-called Case~BB RLO \citep{tlk12}. This post-CE mass-transfer
phase is a result of the naked helium star (the stripped core of the original
IMXB donor star) filling its Roche~lobe when it expands to become a giant
during helium shell burning. Hence, for the purpose of understanding the
recycling of \psr we only have to consider this epoch of evolution in detail. A
complete overview the full progenitor evolution of the system is illustrated in
Fig.~\ref{fig:cartoon}.

\subsection{Calculations of Case~BB RLO leading to Recycling of \psr}
\label{subsec:CaseBB} 

\begin{figure} 
    \centering
    \includegraphics[width=0.80\columnwidth,angle=-90]{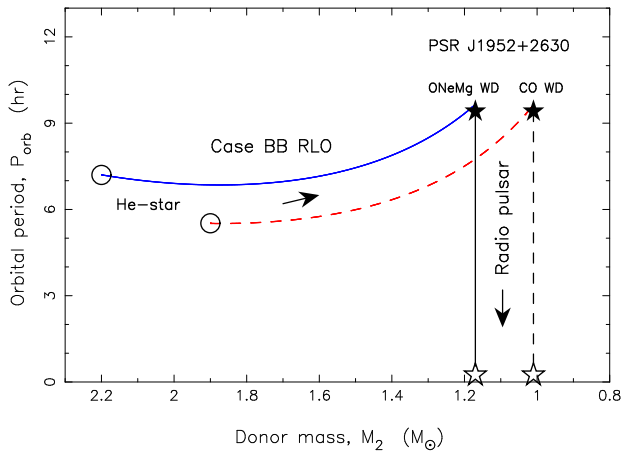} 
    \caption{
        Progenitor evolution of the \psr system in the ($M_2,P_{\rm orb}$)--plane
        during mass transfer via Case~BB RLO.  The blue solid line is the evolutionary
        track for a $2.2\;M_{\odot}$ helium star leaving a $1.17\;M_{\odot}$ ONeMg~WD
        remnant.  The red dashed line is for a $1.9\;M_{\odot}$ helium star leaving a
        $1.02\;M_{\odot}$ CO~WD remnant.  The open circles show the location of the
        helium star donors at the onset of the RLO.  The solid stars indicate the
        termination of the RLO when the radio pulsar turns on. In about 3.4~Gyr the
        system will merge (open stars). 
        \label{fig:M2_Porb}}
\end{figure}

Binary evolution for NS--massive~WD systems have been studied using detailed
calculations of the Case~BB RLO we applied the Langer stellar evolution code
\citep[e.g.][]{tlk11,tlk12}. However, none of the previously computed models
produced systems sufficiently similar to that of \psr. We used the same
code to study what progenitor systems can result in \psr-like binaries, the
Case~BB RLO of these systems, and the nature of the WD companion.

The masses of the WD and the NS ($M_{\rm WD}$ and
$M_{\rm NS}$, respectively) are not known from timing at this stage (see
Sec.~\ref{subsec:timingconstraints}).  In order to limit the number of trial
computations, we assumed $M_{\rm NS}=1.35\;M_{\odot}$ and performed various
calculations with different values of initial orbital period and initial mass
of the helium donor star, $M_2$.

The young age of \psr implies that $P_{\rm orb}$ (now 9.4~hours) has
not changed much by gravitational wave radiation since the termination of the
mass transfer (it was at most $\sim$9.6~hours; see Sec.~\ref{sec:discussion}).
Therefore we only select progenitor solutions of our modelling that have similar
orbital periods. We also impose the criterion $M_{\rm
WD}=$0.93--1.48$\;M_{\odot}$, to be consistent with the minimum companion mass
derived from our timing solution.

Two solutions satisfying our selection criteria are shown in
Fig.~\ref{fig:M2_Porb}.  The first solution (blue solid line) is for a
$2.2\;M_{\odot}$ helium star leading to formation of a $1.17\;M_{\odot}$
ONeMg~WD. The second solution (red dashed line) is for a $1.9\;M_{\odot}$
helium star leading to a $1.02\;M_{\odot}$ CO~WD.  In both cases we assumed a
helium star metallicity of $Z=0.02$ (solar metallicity), a reasonable assumption given that their $\sim$6--7 $M_{\odot}$ progenitors had short lifetimes of $\sim$100~Myr and thus belong to Galactic Population~I stars. 
% Whereas the first solution is somewhat more probable in terms of its WD mass
% (being closer to the median value given the unconstrained orbital inclination
% angle, $i$) 
The second solution predicts a shorter pulsar spin period at the termination of
accretion, and therefore imposes  a less strict limit on the mass-accretion
efficiency; see Sec.~\ref{subsec:age-spin} for a discussion.

Unfortunately, given the current timing data it is not possible to place
sufficiently stringent constraints on the binary system to be used to
select either of the two simulated scenarios as the actual evolution of the
binary (see Sec.~\ref{subsec:timingconstraints}, and Fig.~\ref{fig:m2cosi}).

For the remainder of Sec.~\ref{sec:evolution} we will consider only the ONeMg~WD solution to our modelling. In particular, we will highlight some of the more interesting characteristics of the WD.

\begin{figure} 
    \centering
    \includegraphics[width=0.80\columnwidth,angle=-90]{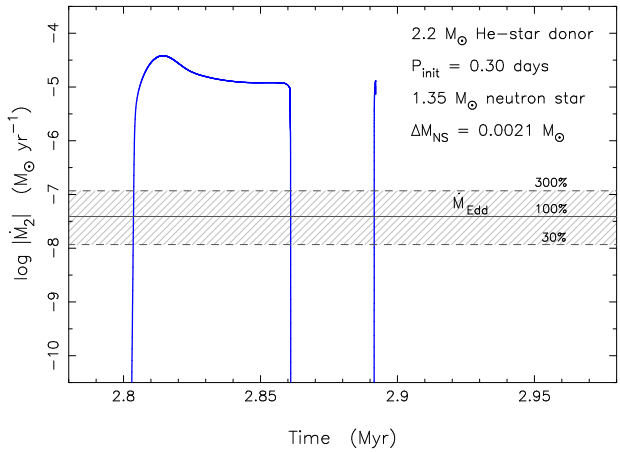} 
    \caption{Mass-transfer rate as a function of stellar age for the progenitor
    evolution plotted in Fig.~\ref{fig:M2_Porb} leading to an ONeMg~WD.  The
    initial configuration is a $2.2\;M_{\odot}$ helium star orbiting a
    $1.35\;M_{\odot}$ NS with $P_{\rm orb}=0.30$~d.  The Case~BB RLO
    lasts for about $60\,000\;{\rm yr}$, then terminates for about
    $30\,000\;{\rm yr}$ until a final vigorous helium shell flash is launched
    (the spike). The mass-transfer rate is seen to be highly super-Eddington
    ($\sim\!10^3\;\dot{M}_{\rm Edd}$).  The horizontal lines mark different
    values for the accretion efficiency in units of the Eddington accretion
    rate, $\dot{M}_{\rm Edd}$.  Depending on the exact value of $\dot{M}_{\rm
    Edd}$, and the accretion efficiency, the NS accretes $0.7-6.4\times
    10^{-3}\;M_{\odot}$, which is sufficient to recycle \psr\ -- see text.
    \label{fig:Mdot}}
\end{figure}

The mass-transfer rate, $|\dot{M}_2|$, for the solution leading to the
$1.17\;M_{\odot}$ ONeMg~WD is shown in Fig.~\ref{fig:Mdot} as a function of
time.  The duration of the Case~BB RLO is seen to last for about $\Delta t
=60\;{\rm kyr}$, which causes the NS to accrete an amount $\Delta M_{\rm
NS}\approx 0.7$--$6.4\times 10^{-3}\;M_{\odot}$, depending on the assumed
accretion efficiency and the exact value of the Eddington accretion limit,
$\dot{M}_{\rm Edd}$. Here we assumed $\dot{M}_{\rm Edd}=3.9\times
10^{-8}\;M_{\odot}\,{\rm yr}^{-1}$ \citep[a typical value for accretion of helium
rich matter,][]{bh91} and allowed for the actual accretion rate to be somewhere in the
interval 30--300\% of this value. This is to account for the fact that the
value of $\dot{M}_{\rm Edd}$ is derived under idealized assumptions of
spherical symmetry, steady-state accretion, Thomson scattering opacity and
Newtonian gravity.  As we shall see, the accretion efficiency, and thus $\Delta
M_{\rm NS}$, is important for the spin period obtained by the NS during its
spin-up phase.

The mass-transfer rate from the helium star is highly super-Eddington
($|\dot{M}_2|\sim 10^{3}\;\dot{M}_{\rm Edd}$). The excess material (99.9\%) is
assumed to be ejected from the vicinity of the NS, in the form of a disc wind
or a jet, with the specific orbital angular momentum of the NS following the
so-called isotropic re-emission model \citep[see][and references therein]{tvs00}.

\subsection{Detailed WD structure}
\label{subsec:WDstructure} 
\begin{figure}
    \centering 
    \includegraphics[width=1.05\columnwidth,height=0.7\columnwidth]{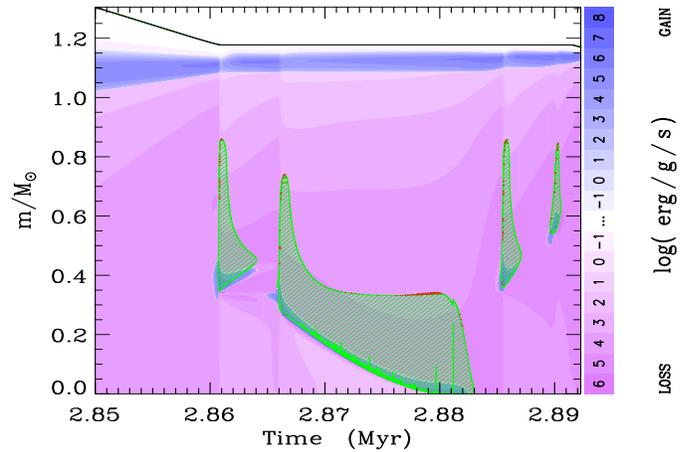} 
    \caption{ The Kippenhahn diagram showing the formation of an ONeMg~WD
    companion to \psr.  The plot shows cross-sections of the progenitor star in
    mass-coordinates from the centre to the surface, along the y-axis, as a
    function of stellar age (since the helium star ZAMS) on the x-axis.  Only
    the last $42\;{\rm kyr}$ of our calculations are plotted.  The Case~BB
    RLO is terminated at time $t=2.86\;{\rm Myr}$ when the progenitor star has
    reduced its mass to $1.17\;M_{\odot}$.  The green hatched areas denote
    zones with convection.  The intensity of the blue/purple colour indicates
    the net energy-production rate; the helium burning shell near the surface
    is clearly seen at $m/M_{\odot}\simeq 1.1$ as well the off-centred carbon
    ignition in shells, starting at $m/M_{\odot}\simeq 0.4$, defining the
    subsequent inner boundaries of the convection zones. The mixing of elements
    due to convection expands the ONeMg core out to a mass coordinate of about
    $m/M_{\odot}\simeq 0.85$ (see Fig.~\ref{fig:abundances}).  Energy losses due
    to neutrino emission are quite dominant outside of the nuclear burning
    shells. \label{fig:kippenhahn}}
\end{figure}

The calculated interior structure and evolution of a likely progenitor of \psr,
a $2.2\;M_{\odot}$ helium star which undergoes Case~BB RLO and leaves behind an
ONeMg~WD, is illustrated in the ``Kippenhahn diagram'' \citep{kw90} in
Fig.~\ref{fig:kippenhahn}. The plot shows the last $10\;{\rm kyr}$ of the
mass-transfer phase ($t=2.85$--$2.86\;{\rm Myr}$), followed by $32\;{\rm kyr}$ of
evolution ($t=2.860$--$2.892\;{\rm Myr}$) during which carbon is ignited in  the
detached donor star. 

In our modelling of the companion star, there are four instances of off-centred
carbon burning shells. These shells are the four blue regions underneath the green-hatched
convection zones in Fig.~\ref{fig:kippenhahn}. The ignition points are
off-centre because these surrounding layers are hotter than the interior due
to more efficient neutrino cooling in the higher-density inner core. The
maximum temperature is near a mass coordinate of $m/M_{\odot} \simeq 0.4$.

\begin{figure} 
    \centering
    \includegraphics[width=0.80\columnwidth,angle=-90]{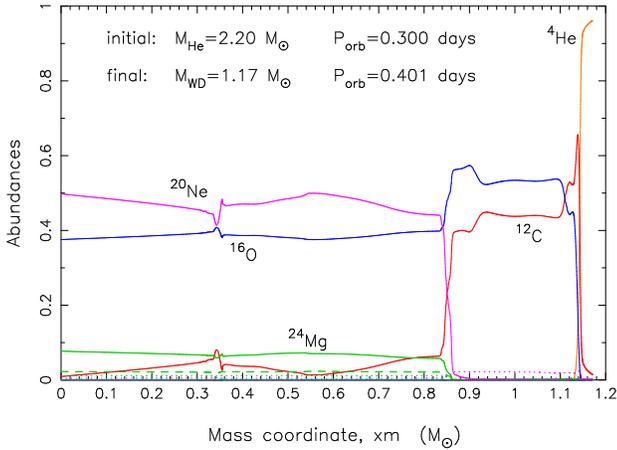}
    \caption{ The chemical abundance structure of the ONeMg~WD remnant (from
    our last calculated model at $t=2.892\;{\rm Myr}$) of the Case~BB RLO
    calculation shown in Figs.~\ref{fig:M2_Porb}--\ref{fig:kippenhahn},
    one of the plausible solutions for \psr's companion found in our modelling.  This
    $1.17\;M_{\odot}$ WD has a hybrid structure with an ONeMg core enclosed by
    a CO mantle and a tiny ($0.027\;M_{\odot}$) surface layer of helium. 
    \label{fig:abundances}}
\end{figure}

The second carbon-burning shell penetrates to the centre of the proto-WD.
However, at no point in the modelled evolution of the companion do the carbon
burning shells, or the associated convection zones on top of these shells,
reach the surface layers of the proto-WD. Therefore, the resulting WD structure
is a hybrid, with a large ONeMg core engulfed by a thick CO mantle.  The
chemical abundance profile of the WD companion at the end of our modelling
($t=2.892\;{\rm Myr}$) is demonstrated in Fig.~\ref{fig:abundances}. Notice
the tiny layer ($2.7\times 10^{-2}\;M_{\odot})$ of helium at the surface which
gives rise to a vigorous helium shell flash at $t=2.892\;{\rm Myr}$. This shell
flash can also be seen in Figs.~\ref{fig:Mdot} and \ref{fig:kippenhahn} and
gives rise to numerical problems for our code. We therefore end our
calculations without resolving this flash. However, since the NS is only
expected to accrete of the order $\sim\!10^{-5}\;M_{\odot}$ as a result of this
flash (based on modelling of similar binaries where we managed to
calculate through such a helium shell flash), its impact on the final binary
and spin parameters will be completely negligible.

As far as we are aware, this is the first presentation in the
literature of detailed calculations leading to an ONeMg~WD orbiting a recycled
pulsar.

\section{Discussion}
\label{sec:discussion}
The future and past spin evolution of \psr can be computed from the measured
values of $P$ and $\dot{P}$ and assuming a (constant) braking index, $n$,
which is defined by $\dot{\Omega}=-K\Omega ^{n}$, where $\Omega =2\pi/P$ and
$K$ is a scaling factor \citep{mt77}. The resulting future and the past spin
evolution of \psr are shown in Fig.~\ref{fig:spin_evol}, top and bottom panel,
respectively. 

Given our modelling, the true age of \psr is \lapp 150$\;{\rm Myr}$. If a
cooling age of the WD companion of a pulsar in a similar system could be
accurately determined (for which the WD mass is needed), it would be possible
to constrain the braking index of a recycled pulsar. Although this is not
likely possible in the case of \psr due to its large (DM) distance of 9.6~kpc,
it may be feasible for other systems in the future.

The orbital decay due to gravitational wave radiation has also been computed
(see Fig.~\ref{fig:spin_evol}, top). It is evident that $P_{\rm orb}$ has
hardly decayed since the formation of \psr. Sec.~\ref{subsec:age-spin} describes
the implication for our binary stellar evolution modelling.

\subsection{On the age and spin evolution of \psr}
\label{subsec:age-spin}
Our modelling of binary stellar evolution and accretion physics provides initial
conditions for the spin of \psr, which must be consistent with current
measurements. In particular, the spin period of \psr predicted at the
termination of accretion must be smaller than the observed spin period.

\begin{figure} 
    \centering
    \includegraphics[width=1.00\columnwidth,angle=0]{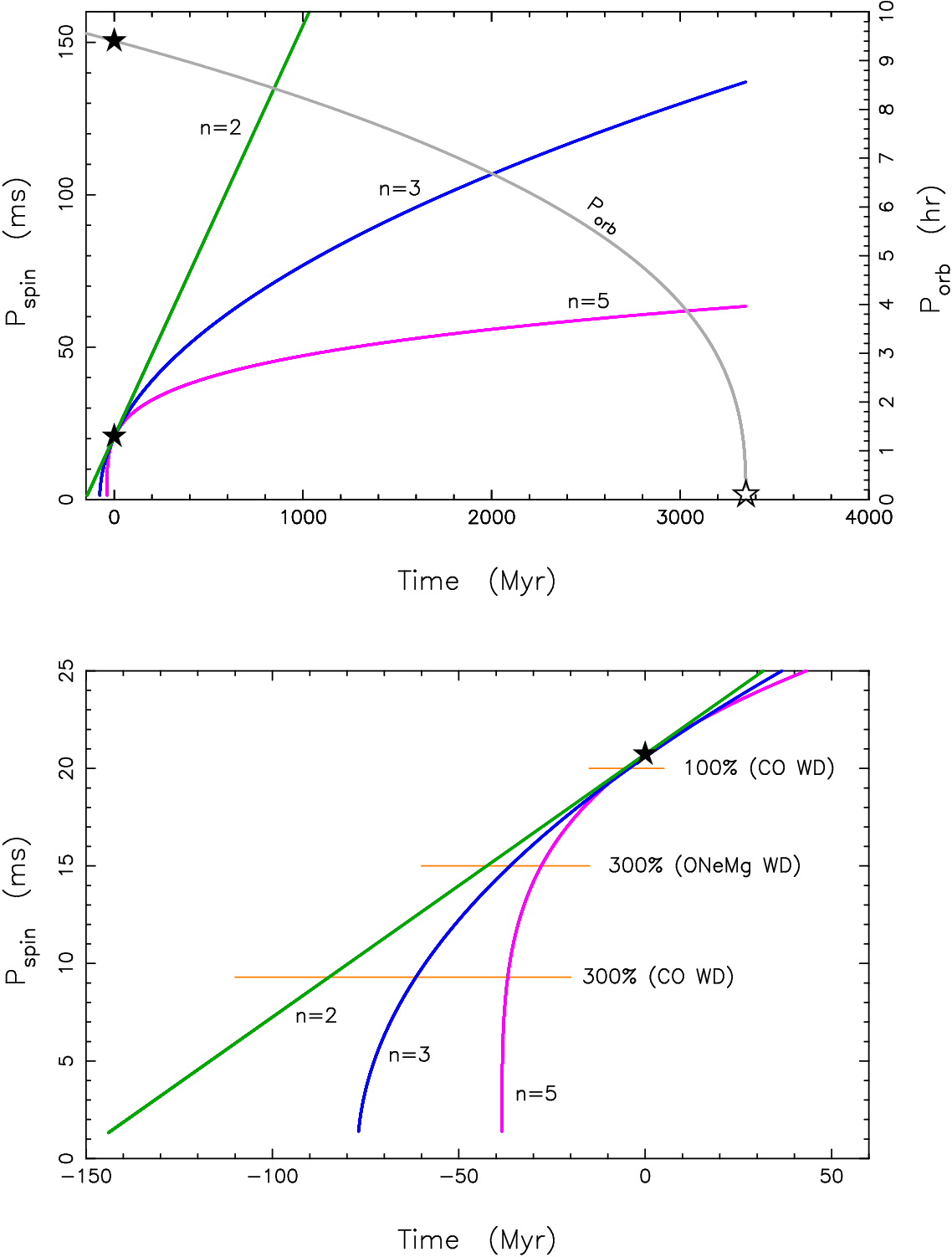} 
    \caption{ The future (top) and the past (bottom) spin evolution of \psr
    for different values of the braking index, $n$. The location of \psr at
    present is marked by a solid star. In the top panel, the grey curve shows
    the calculated orbital decay due to gravitational wave radiation until the
    system merges in about 3.4~Gyr (marked by an unfilled star). The lower panel is
    a zoom-in on the past spin evolution. Depending on $n$, the WD companion
    mass and the accretion efficiency of the NS during Case~BB RLO, the pulsar
    could have been spun~up to the initial spin periods indicated
    by the orange horizontal lines -- see text for a discussion.
    \label{fig:spin_evol}}
\end{figure}

The minimum equilibrium spin period can be estimated given the amount of mass
accreted by the NS, $\Delta M_{\rm NS}$. This quantity depends on our binary
evolution models and the assumed accretion efficiency (in units of
$\dot{M}_{\rm Edd}$, see Fig.~\ref{fig:Mdot}). Following  Eqn. 14 of
\citet{tlk12}, 

\begin{equation} 
  P_{\rm ms} = \frac{(M_{\rm NS}/M_{\odot})^{1/4}}{(\Delta
               M_{\rm NS}/0.22\;M_{\odot})^{3/4}} , \label{eq:eq_spin} 
\end{equation}

\noindent where $P_{\rm ms}$ is the equilibrium spin period in units of ms, we estimate
the minimum equilibrium spin period of \psr. These results are tabulated in
Table~\ref{table:CaseBB}, and compared with \psr's current spin period in
Fig.~\ref{fig:spin_evol} (bottom). It is interesting to notice that we only
obtain solutions for an accretion efficiency of 100\% or 300\% of $\dot{M}_{\rm
Edd}$\footnote{Solutions requiring larger-than-$\dot{M}_{\rm Edd}$ accretion
efficiencies are in fact physically viable because assumptions made during the
calculation of $\dot{M}_{\rm Edd}$ mean it is only a rough measure of the true
limiting accretion rate.}.  If the accretion efficiency is smaller, the
equilibrium spin period becomes larger that the present spin period of
$P=20.7\;{\rm ms}$, which is impossible.

The reason for the possibility of a lower initial spin period in case \psr has
a CO~WD companion, is simply that the helium star progenitor of a CO~WD has a
lower mass and therefore evolves on a longer time-scale, thereby increasing
$\Delta M_{\rm NS}$.  Hence, a cooling age estimate of the WD companion could,
in principle, also help constrain the accretion physics, because it puts
limitations on the possible values of the initial spin period. 

This is the first time an accretion efficiency has been constrained for a
recycled pulsar which evolved via Case~BB RLO. In contrast, the accretion
efficiency of millisecond pulsars formed in low-mass X-ray binaries (LMXBs) has
been shown to be much lower, about 30\% in some cases
\citep{ts99,jhb+05,akk+12}. The reason for this difference in accretion
efficiencies may be related to the extremely high mass-transfer rates during
Case~BB RLO which could influence the accretion flow geometry and thus
$\dot{M}_{\rm Edd}$. Furthermore, accretion disc instabilities
\citep{las01,cfd12}, which act to decrease the accretion efficiency in LMXBs,
do not operate in Case~BB RLO binaries, due to the high value of $|\dot{M}_2|$. 

\subsubsection{Evolution in the $P$-$\dot{P}$ diagram}
\label{subsubsec:isochrones} 
\begin{figure*} 
    \centering
    \includegraphics[width=1.25\columnwidth,angle=-90]{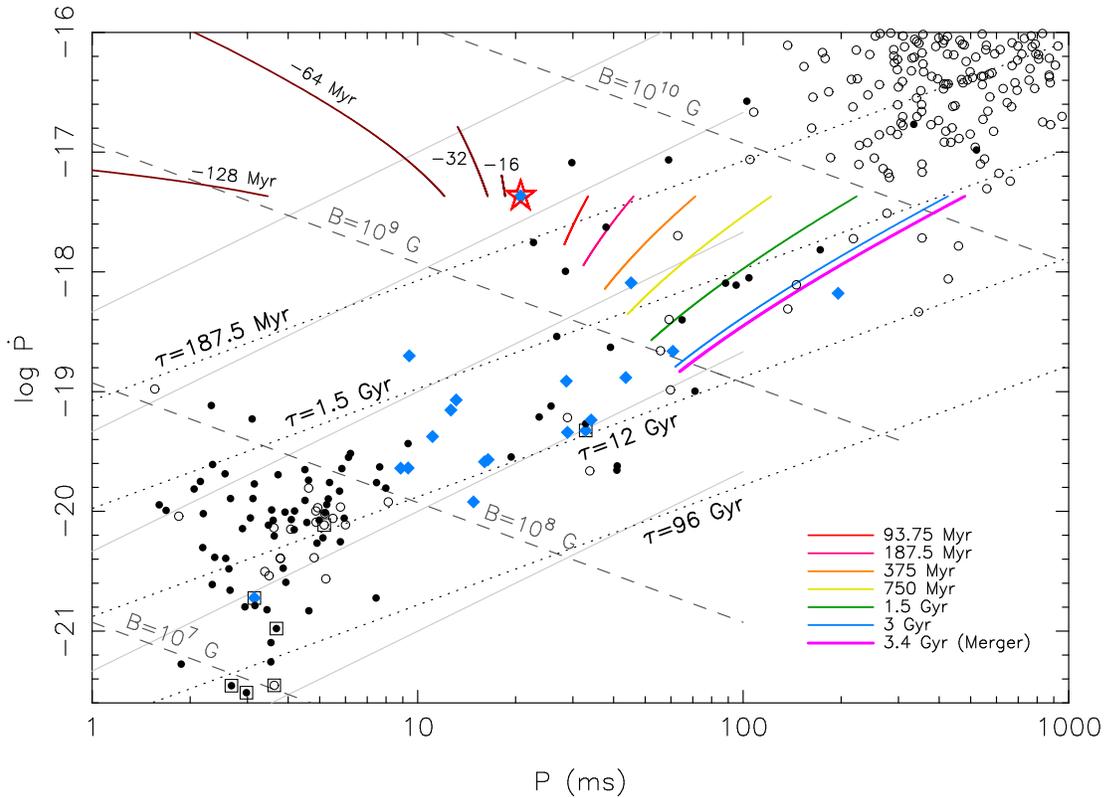} 
    \caption{ Isochrones of past (brown colour) and future evolution (rainbow
    colours) of \psr in the $P$-$\dot{P}$ diagram.  The present location is
    marked by the open red star.  All isochrones were calculated for braking
    indices in the interval $2\le n\le 5$.  Also plotted are inferred constant
    values of $B$-fields (dashed lines) and characteristic ages, $\tau$ (dotted
    lines).  The thin grey lines are spin-up lines with $\dot{M}/\dot{M}_{\rm
    Edd}=1,\,10^{-1},\,10^{-2},\,10^{-3}$ and $10^{-4}$ (top to bottom),
    assuming a pulsar mass of $1.4\,M_{\odot}$.  Observational data of the plotted
    Galactic field pulsars are taken from the ATNF Pulsar Catalogue in
    March~2013.  Binary pulsars are marked with solid circles and isolated
    pulsars are marked with open circles. For further explanations of the
    calculations, and corrections to $\dot{P}$ values see \citet{tlk12}.
    Binary pulsars with a massive (CO or ONeMg)~WD companion are marked with a
    blue diamond. The past spin evolution of \psr is particularly interesting
    as it constrains both the binary evolution and the recycling process
    leading to its formation -- see text.
%            -- see text for discussions.
    \label{fig:isochrones}} 
\end{figure*}

By integrating the pulsar spin deceleration equation: $\dot{\Omega}=-K\Omega
^{n}$, assuming a constant braking index, $n$ we obtain isochrones.  The
kinematic solution at time $t$ (positive in the future, negative in the past)
is given by: 
\begin{equation}
\label{eq:spindown}
      P = P_0 \left[ 1 + (n-1) \frac{{\dot P}_0}{P_0}\,t \right]^{1/(n-1)}
\end{equation}

\begin{equation}
\label{eq:spindown_dot}
 \dot P = {\dot P}_0 \left( \frac{P}{P_0} \right)^{2-n}, 
\end{equation}

\noindent where $P_0 = 20.7\;{\rm ms}$ and ${\dot P}_0 = 4.27 \times 10^{-18}
\rm s\,s^{-1}$ are approximately the present-day values of the spin period and
its derivative. The past and future spin evolution of \psr in the $P$-$\dot{P}$
diagram are plotted in Fig.~\ref{fig:isochrones}. The isochrones are calculated
using Eqns.~\ref{eq:spindown} and \ref{eq:spindown_dot} where $n$ varies from 2
to 5, for different fixed values of $t$ in the future (rainbow colours) and
past (brown).  For each isochrone, the time is given by the well-known
expression:

\begin{equation}
\label{eq:trueage}
  t=\frac{P}{(n-1)\dot{P}}\left[1-\left(\frac{P_0}{P}\right)^{n-1}\right].
\end{equation}

%We summarize the past and the future spin evolution in the $P$-$\dot{P}$ diagram
%in Fig.~\ref{fig:isochrones}.  The isochrones are obtained by integrating the
%pulsar spin evolution using the following expression for the true age of the
%pulsar, $t$: 
%\begin{equation} t=\frac{P}{(n-1)\dot{P}}\left[
%1-\left(\frac{P_0}{P}\right)^{n-1}\right], \label{eq:trueage} 
%\end{equation}
%where $P_0$ is the current spin period at time $t=0$. For the braking index we
%assumed $2\le n\le5$.  The brown isochrones show the potential past location
%($t<0$) of \psr. However, only small values of $n$ yield solutions for an
%increasing past age of the pulsar -- cf. Fig.~\ref{fig:spin_evol}.
%Furthermore, these solutions are purely based on rotational kinematics.  As
These solutions, however, are purely based on rotational kinematics.  As
already discussed above, one must take the constraints obtained from binary
evolution and accretion physics into account.  Therefore, if the Case~BB RLO is
not able to spin~up the pulsar to a value smaller than, for example, 15~ms,
then the true age of \psr cannot be much more than 40~Myr for all values of
$2\le n\le 5$. For closer binary systems similar to \psr where a
cooling age determination of the WD is possible, such a measurement could be
useful for constraining the birth period of the pulsar, as well as the system's
accretion efficiency. 

% From an email to Bruce Allen on May 22, 2013
%
% A 66 Myr old ONeMg WD has an absolute magnitude (i.e. magnitude if the object
% were at a distance of 10 pc) of ~10. Scaling this to the DM-distance of
% J1952+2630 gives an apparent magnitude of 25. However, the system is in the
% Galactic plane (lat=-0.4 deg). The extinction along the line-of-sight would
% cause the WD to appear up to ~17 magnitudes fainter. The faintest objects
% seen with 8-m class ground-based telescopes is ~27.

Also shown in Fig.~\ref{fig:isochrones} is the future spin evolution of \psr
until the system merges in about 3.4~Gyr.  It is interesting to notice that a
system like \psr should be observable as a radio pulsar binary until it
merges, suggesting the existence of similar NS--massive~WD binaries with
much shorter orbital periods, to which PALFA is
sensitive \citep{akc+13,laz13}.  The unique location of \psr with respect to
other known recycled pulsars with a massive WD companion (marked with blue
diamonds) is also clear from this figure. It is seen that only three other
systems may share a past location in the $P$-$\dot{P}$ diagram similar to that of
\psr (if $2\le n\le 5$). This could suggest that such surviving post-CE systems are 
often formed with small values of $P_{\rm orb}$ which cause them to merge
rapidly -- either during the Case~BB RLO or shortly thereafter due to
gravitational wave radiation. 

%%%%%%%%%%%%%%%%%%%%%%%%%% TT end %%%%%%%%%%%%%%%%%%%%%%%%%%%%%%%%%%%%%%

\section{Future Prospects for \psr}
\label{sec:future}

We now investigate the future use of \psr as a gravitational
laboratory. Looking at Table~\ref{tab:timing}, we can see that the
eccentricity, $e$, and longitude of periastron, $\omega$, can be measured quite
precisely, in the latter case to within 1.2$^{\circ}$, despite the small absolute
value of the eccentricity, 4.1(1) $\times \, 10^{-5}$.  If we assume a mass of
1.35 $M_{\odot}$ for the pulsar and 1.1 $M_{\odot}$ for the WD, then general
relativity predicts that $\omega$ should increase at a rate
$\dot{\omega}=1.72^\circ \,\rm yr^{-1}$, which given the precision of
$\omega$ implies that the effect should be detectable in the next few years. Measuring it will
eventually give us an estimate of the total mass of the system \citep{wt81}.

Furthermore, thanks to \psr's rather short orbital period, the shortest among
recycled pulsar--massive WD systems, the rate of gravitational wave
emission is much higher than for any other such system. This emission will
cause the orbit to decay. For the same assumptions as above, the orbital period
should change at a rate $\dot{P}_{b{\rm, pred}}=-1.14 \times 10^{-13}\, \rm s\,
s^{-1}$. As previously mentioned, this will cause the system to merge within
about 3.4 Gyr.

The orbital decay due to the emission of gravitational waves is not measurable
at present but it should be detectable in the near future. In order to verify
this, we made simulations of future timing of this pulsar that assume similar
timing precision as at present (a single 18-$\mu$s TOA every 15 minutes for each of
two 300-MHz bands centred around 1400 MHz). We assume two campaigns in 2015
and 2020, where the full orbit is sampled a total of 8 times, spread evenly for
each of those years. The total observing time of 72 hours
for each of those years is a realistic target.  These simulations indicate that
by the year 2020 $\dot{P}_b$ should be detectable with 12-$\sigma$
significance and $\dot{\omega}$ to about 9-$\sigma$ significance.

\begin{figure*} 
    \centering
    \includegraphics[width=1.8\columnwidth,angle=0]{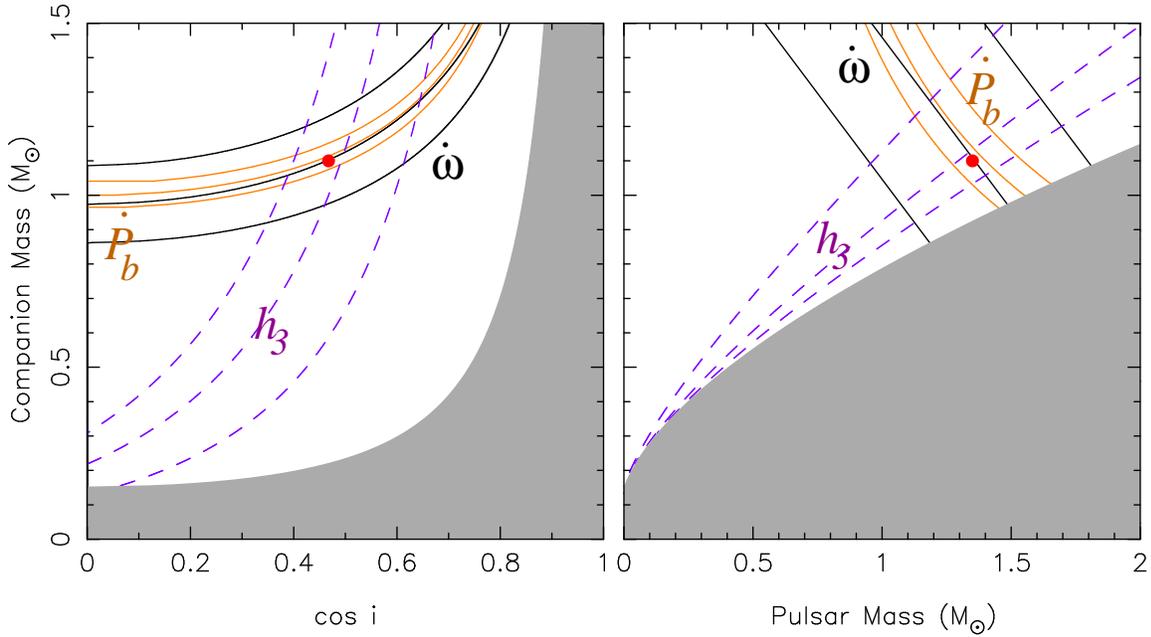} 
    \caption{Constraints on system masses and orbital inclination from
     simulated radio timing of \psr.
     Each triplet of curves corresponds to the most likely value and standard 
     deviations of the respective parameters; the region limited by it is the
     ``1-$\sigma$'' band for that parameter. The masses and inclination
     used in the simulation are indicated by the red points.  {\bf Left}: $\cos i$--$M_{\rm WD}$
     plane. The grey region is excluded by the condition $M_{\rm NS} >$ 0. 
     {\bf Right}: $M_{\rm NS}$--$M_{\rm WD}$ plane. The grey region is 
     excluded by the condition $\sin i \leq$ 1. 
    \label{fig:mass_mass_sim}} 
\end{figure*}

The resulting constraints on the masses of the components and system
inclination are depicted graphically in Fig.~\ref{fig:mass_mass_sim}. We plot
the 1-$\sigma$ bands allowed by the `measurement' of a particular parameter.
The figure shows several interesting features. The first is that even with
these measurements of $\dot{P}_b$ and $\dot{\omega}$, it is not
possible to determine the two masses accurately from them, given the way their
1-$\sigma$ uncertainty bands intersect in the mass-mass diagram.  The
1-$\sigma$ band of $h_3$ (the orthometric amplitude of the Shapiro delay, see
Freire \& Wex 2010\nocite{fw10}) intersects the others at a rather sharp angle
and can in principle be used to determine the masses more accurately.  However,
at the moment it is not clear whether $h_3$ is precisely measurable.  A more
massive companion in a less inclined orbit yields more pessimistic
expectations, as depicted in Fig.~\ref{fig:mass_mass_sim}.

Fortunately, very precise masses are not required to perform a test of general
relativity using $\dot{\omega}$ and $\dot{P}_b$. As illustrated in
Fig.~\ref{fig:mass_mass_sim}, the bands allowed by $\dot{\omega}$ and
$\dot{P}_b$ are nearly parallel. As the precision of these measurements
improves general relativity is tested by the requirement that the two bands
still overlap. Fig.~\ref{fig:mass_mass_sim} also shows that even if we could
determine $h_3$ precisely, the precision of this
$\dot{\omega}$-$h_3$-$\dot{P}_b$ test will be limited by the precision of the
measurement of $\dot{\omega}$, because its uncertainty only
decreases with $T^{-3/2}$ (where $T$ is the timing baseline), while for
$\dot{P}_{b}$ the measurement uncertainty decreases with $T^{-5/2}$.

Eventually, though, the uncertainty of the measurement of $\dot{\omega}$
will become very small. At that stage the precision of this test will be
limited by the precision of $\dot{P}_b$, which is limited by the lack of
precise knowledge of the kinematic contributions \citep{dt91}:
\begin{equation}
        \left(  \frac{\dot{P}_b}{P_b} \right) = \left(  \frac{\dot{P}_b}{P_b} \right)_{\rm obs} - \frac{\mu^2 d}{c} - \frac{a(d)}{c},
\end{equation}
where the subscript ``obs'' indicated the observed quantity,
$\mu$ is the total proper motion, $d$ is the distance to the
pulsar and $a(d)$ is the (distance dependent) difference between the
Galactic acceleration of the system and that of the Solar System
Barycentre (SSB), projected along the direction from the SSB to the
system. At present these cannot be estimated because the proper motion
has not yet been measured precisely. However, they likely represent the
ultimate constraint to the precision of this GR test, as in the case of
PSR~B1913+16 \citep{wnt10}. For that reason, we now estimate
the magnitude of these kinematic effects.

If we assume that the final proper motion is of the same order of magnitude as
what is observed now ($\sim\!6\;{\rm mas}\,{\rm yr}^{-1}$), then at
the assumed DM-distance of 9.6~kpc the kinematic contribution to $\dot{P}_b$
will be about $3 \times 10^{-14}\,\rm s \,s^{-1}$.  This is four
times smaller than $\dot{P}_{b{\rm, pred}}$, as defined above, which means that
if we can't determine the distance accurately, then the real value for the
intrinsic orbital decay, $\dot{P}_{b, \rm int}$, cannot be measured with a
relative precision better than about 25~\%.  Things improve if the
proper motion turns out to be smaller.

Is such a measurement useful? The surprising answer is that it is likely to be
so. For pulsar-WD systems, alternative theories of gravity, like Scalar-Tensor
theories \citep[see][and references therein]{de98} predict the emission of
dipolar gravitational waves, which would result in an increased rate of orbital
decay.  In the case of PSR~J1738+0333, the orbital decay for that system was
measured only with a significance of 8-$\sigma$. However, the absolute
difference between the $\dot{P}_b$ predicted by GR and the observed value is so
small that it introduces the most stringent constraints ever on these gravity
theories \citep{fwe+12}. The implication is that for \psr, one should be able
to derive similarly low limits. However, if the proper motion is
significantly smaller, and/or if we are able to determine the distance
independently, then this system can provide a much more stringent test of
alternative theories of gravity.  The main reason for this is that the limiting
factor of the PSR J1738+0333 test is the limited precision in the measurement
of the component masses \citep{akk+12} which would not be an issue for \psr,
given the tighter constraints on the total mass that will eventually be derived
from $\dot{\omega}$.

\section{Conclusions}
\label{sec:conclusion}
We have presented phase-coherent timing of \psr. Our timing model includes
precise determinations of parameters describing the pulsar spin-down,
astrometry and binary motion. No post-Keplerian orbital parameters are
required.  However, detailed modelling suggests the current pulsar--massive WD
binary system evolved via post-CE Case~BB RLO with an accretion efficiency
which exceeded the Eddington limit by a factor of 1--3. We presented, for the
first time, a detailed chemical abundance structure of an ONeMg~WD orbiting a
pulsar.

By projecting \psr's orbital evolution into the future we estimate it will
merge with its WD companion in $\sim$3.4~Gyr due to the orbital decay from
gravitational wave emission. Unfortunately, \psr is too distant to make a
detection of the cooling age of the pulsar's WD companion. In the case of the discovery
of a less distant analog of \psr such a measurement could make it possible to
constrain the braking index of the recycled pulsar, and/or the accretion
efficiency during the Case~BB RLO-phase. Also, timing observations over the
next 10 years will result in the detection of the advance of periastron, and
the orbital decay, enabling a test of general relativity. Finally,
additional timing may also further elucidate the nature of the companion, and
will permit \psr to be used to perform gravitational tests.
\\

The Arecibo Observatory is operated by SRI International under a
cooperative agreement with the National Science Foundation (AST-1100968),
and in alliance with Ana G. M\'endez-Universidad Metropolitana, and the
Universities Space Research Association. Einstein@Home is supported by
the Max Planck Society and US National Science Foundation (NSF) grants 1104902,
1104617, 1105572, and 1148523.
\\
\noindent We thank Norbert Langer for discussions, as well as Jason
Hessels and Aristeidis Noutsos for helpful comments.  \\
PL acknowledges the support of IMPRS Bonn/Cologne and NSERC PGS-D.  TMT
gratefully acknowledges support and hospitality from the Argelander-Insitut
f\"ur Astronomie, Universit\"at Bonn and the Max-Planck-Institut f\"ur
Radioastronomie.  BK acknowledges the support of the Max Planck Society. PCCF
gratefully acknowledges financial support by the European Research Council for
the ERC Consolidator Grant BEACON under contract no. 279702. BK acknowledges
the support of the Max Planck Society. VMK was supported by an NSERC Discovery
Grant, the Canadian Institute for Advanced Research, a Canada Research Chair,
Fonds de Recherche Nature et Technologies, and the Lorne Trottier Chair in
Astrophysics. 
%JWTH acknowledges support from the Netherlands Foundation for
%Scientific Research (NWO). 
Pulsar research at UBC is supported by an NSERC Discovery Grant and Discovery
Accelerator Supplement.
\\

%\bibliographystyle{mn2e}
%\bibliography{J1952_refs}

\clearpage

\begin{table}
\caption{Fitted and derived parameters for PSR~J1952+2630.}
\label{tab:timing}
\begin{tabular}{lc}
        \hline
        Parameter & Value$^{a}$\\
        \hline
        \multicolumn{2}{c}{\textit{General Information}} \\
        \hline
        MJD Range & 55407 -- 56219 \\
        Number of TOAs & 418 \\
        Weighted RMS of Timing Residuals ($\mu$s) & 19 \\
        Reduced-$\chi^2$ value$^{b}$ & 1.03 \\
        MJD of Period Determination & 55813 \\
        Binary Model Used & ELL1 \\
        \hline
        \multicolumn{2}{c}{\textit{Fitted Parameters}} \\
        \hline
        R.A., $\alpha$ (J2000) & 19:52:36.8401(1) \\
        Dec., $\delta$ (J2000) & 26:30:28.073(2) \\
        Proper motion in R.A., $\mu_\alpha$ (mas/yr) & -6(2) \\
        Proper motion in Dec., $\mu_\delta$ (mas/yr) & 0(3) \\
        Spin Frequency, $\nu$ (Hz) & 48.233774295845(7) \\
        Spin Frequency derivative, $\dot{\nu}$ ($\times 10^{-15}$ Hz/s) & -9.9390(5) \\
        Dispersion Measure, DM (pc~cm$^{-3}$) & 315.338(2) \\
        Projected Semi-Major Axis, $a\,\sin i$ (lt-s) & 2.798196(2) \\
        Orbital Period, $P_b$ (days) & 0.39187863896(7) \\
        Time of Ascending Node, $T_{asc}$ (MJD) & 55812.89716459(4) \\
        $\epsilon_1$ & -0.000038(1) \\
        $\epsilon_2$ & 0.000015(1) \\
        \hline
        \multicolumn{2}{c}{\textit{Derived Parameters}} \\
        \hline
	Spin Period, (ms) & 20.732360562672(3) \\
        Spin Period Derivative ($\times 10^{-18} \rm s\,s^{-1}$) & 4.2721(2) \\
        Galactic longitude, $l$ ($^{\circ}$) & 63.254 \\
	Galactic latitude, $b$ ($^{\circ}$) & $-$0.376 \\
	Distance (NE2001, kpc) & 9.6 \\
        Orbital Eccentricity, $e$ ($\times 10^{-5}$) & 4.1(1) \\
        Longitude of Periastron, $\omega$ ($^{\circ}$) & 291(2) \\
        Mass Function, $f$ ($M_{\odot}$) & 0.153184(1) \\
        Characteristic Age, $\tau_c = P/(2\dot{P})$ (Myr) & 77 \\
        Inferred Surface Magnetic Field Strength, $B_S$ ($\times 10^{9}$ G) & 9.5 \\
        Spin-down Luminosity, $\dot{E}$ ($\times 10^{35}$ ergs/s) & 0.19 \\
        \hline
\end{tabular}
\medskip
$^{a}$The numbers in parentheses are the 1-$\sigma$, \tempotwo-reported uncertainties on the last digit.\\
$^{b}$The uncertainties of the ALFA and L-wide data sets were individually scaled such that the reduced $\chi^{2}$ of the data sets are 1.
\end{table}

\begin{table}
\caption{Equilibrium spin obtained via Case~BB RLO.}
\label{table:CaseBB}
\begin{tabular}{cccc}
        \hline
        %\multicolumn{5}{c}{$2.2\;M_{\odot}$ helium star $\longrightarrow 1.17\;M_{\odot}$ ONeMg~WD} \\
        %\hline
        Acc. eff. & $\Delta t_{\rm RLO}$ (kyr) & $\Delta M_{\rm NS}$ ($M_{\odot}$) & $P$ (ms)\\
        \hline
        \multicolumn{4}{c}{\textit{$2.2\;M_{\odot}$ He~star $\longrightarrow 1.17\;M_{\odot}$ ONeMg~WD}} \\
        \hline
        30\%   & 57 & $0.7\times 10^{-3}$ & 80\\
        100\%  & 61 & $2.1\times 10^{-3}$ & 35\\
        300\%  & 61 & $6.4\times 10^{-3}$ & 15\\
        \hline
        \multicolumn{4}{c}{\textit{$1.9\;M_{\odot}$ He~star $\longrightarrow 1.02\;M_{\odot}$ CO~WD}} \\
        %\hline
        % & Acc. eff. & $\Delta t_{\rm RLO}/{\rm kyr}$ & $\Delta M_{\rm NS}/M_{\odot}$ & $P_{\rm ms}$\\
        \hline
        30\%   & 113 & $1.5\times 10^{-3}$  & 45\\
        100\%  & 119 & $4.4\times 10^{-3}$  & 20\\
        300\%  & 113 & $12.4\times 10^{-3}$ & 9.3\\
        \hline
\end{tabular}
\end{table}

\end{document}